\documentclass[preprint,12pt]{elsarticle}%
\usepackage{color}
\usepackage{amsmath}
\usepackage{amsfonts}
\usepackage{amssymb}
\usepackage{hyperref}
\usepackage{graphicx}%
\providecommand{\U}[1]{\protect\rule{.1in}{.1in}}

\journal{Physics Letters A}

\begin{document}

\begin{frontmatter}
\title{Device-independent randomness amplification with a single device}
\author[Bra,Brn]{Martin Plesch}
\author[Brn]{Matej Pivoluska}
\address[Bra]{Institute of Physics, Slovak Academy of Sciences, Bratislava, Slovakia}
\address[Brn]{Faculty of Informatics, Masaryk University, Brno, Czech Republic}

\begin{abstract}
Expansion and amplification of weak randomness with untrusted
quantum devices has recently become a very fruitful topic of research.
Here we contribute with a procedure for amplifying a single weak random source
using tri-partite GHZ-type entangled states. If the quality of the
source reaches a fixed threshold $R=\frac{1}{4}\log_{2}(10)$, perfect random bits can be produced. This technique can be
used to extract randomness from sources that can't be extracted neither
classically, nor by existing procedures developed
for Santha--Vazirani sources. Our protocol works with a single fault-free device decomposable
into three non-communicating parts, that is repeatedly reused throughout the amplification process.

\end{abstract}


\end{frontmatter}

\section{Introduction}

Randomness is an invaluable resource in today's computer science. The need for
randomness of very high quality (close to uniformly distributed and
uncorrelated to any other existing data) is evident especially in the field of
cryptography, where malfunctioning random number generators can cause
catastrophic failures and lead to total loss of security
\cite{LenstraHughesAugierEtAl-Ronwaswrong-2012,
HeningerDurumericWustrowEtAl-MiningYourPs-2012,
McInnesPinkas-ImpossibilityofPrivate-1991,
BoudaPivoluskaPleschEtAl-Weakrandomnessseriously-2012,
HuberPawlowski-Weakrandomnessin-2013}.

However, on the classical level no true randomness is available. Production of
pseudo-random sequences is based on assumptions on inaccessibility of certain
information to the adversary, such as thermal noise of semiconductors or
movement of mouse cursor of a computer user. In classical cryptography
different techniques are used to tackle with such limited sources of
randomness. Depending on the available resources, randomness extractors can
either produce almost perfect randomness from a weak source and a short
perfectly random key (so called \emph{seed}), or they can use several
independent weak random sources to produce a shorter, but almost perfect
output (see \cite{Shaltiel-IntroductiontoRandomness-2011} for a survey).
Nevertheless in the most pessimistic adversarial scenario one is unable to
rule out adversary's full knowledge of the underlying processes, because
classical physical theories are deterministic.

With quantum protocols production of random numbers seems to be easy,
thanks to inherent randomness of quantum physics -- measurement in a basis
complementary to the basis in which the states were produced can guarantee a
source of perfect randomness. Thus, if one can trust the devices used for
randomness production, the task is theoretically trivial and experimentally
feasible up to commercial applications \cite{-IDQuantique:-}.

One can, however, go a bit further and ask if the production of random numbers
could be safe not only against an external adversary, but also towards the
supplier of the device itself. The importance of this requirement is
underlined by the experimental complexity and fragility of quantum devices,
which practically prohibits direct testing of processes appearing within the
device. Such security can be indeed achieved by using devices utilizing
quantum states that exhibit super-classical correlation properties, which can
be tested solely by processing input and output data. This check of the
honesty of the devices, which is often performed simultaneously with the
implemented protocol, is referred in a broader scope as device independence.

To design a device independent random number generator, one can use the fact
that states exhibiting super-classical correlation properties exhibit
intrinsic randomness if measured locally. A line of research was devoted to
expansion of free randomness using quantum devices (see e.~g.
\cite{ColbeckKent-Privaterandomnessexpansion-2011,
PironioAc'inMassarEtAl-Randomnumberscertified-2010,
VaziraniVidick-CertifiableQuantumDice-2011,
CoudronYuen-InfiniteRandomnessExpansion-2013,
MillerShi-Robustprotocolssecurely-2014}), which, in the spirit of seeded
randomness extractors, expands the length of preexisting independent random
seed. Recently several researchers attempted to solve the problem of perfect
randomness production, suggesting ways to \emph{amplify} existing weak
randomness with the use of untrusted quantum devices (characterized either
as Santha-Vazirani source \cite{ColbeckRenner-Freerandomnesscan-2012,
GallegoMasanesEtAl-Fullrandomnessfrom-2013,
GrudkaHorodeckiHorodeckiEtAl-Freerandomnessamplification-2013,
MironowiczPawlowski-Amplificationofarbitrarily-2013,BrandaoRamanathanGrudkaEtAl-RobustDevice-IndependentRandomness-2013}
or min-entropy source \cite{ChungShiWu-PhysicalRandomnessExtractors-2014,
BoudaPawlowskiPivoluskaEtAl-Device-independentrandomnessextraction-2014}). In
a related recent work \cite{KohHallSetiawanEtAl-EffectsofReduced-2012,
ThinhSheridanScarani-Belltestswith-2013} authors examine the minimal
properties of random seed output needed to perform Bell tests. Some of these
works consider even more general scenario, in which the adversary is only
restricted by no-signaling. This can also be seen as an attempt to minimize
the assumptions for which independent perfect randomness exists.

Within this paper we contribute to the topic of production of perfect
randomness with the use of untrusted quantum devices and a single weak source
of available randomness. We will show that production of nearly perfect random
numbers is possible in a very simple and experimentally feasible scenario with
a rather weak demand on the weak random source. We utilize three-partite
GHZ-type entanglement, which leads to a possibility of distinguishing between
classical and quantum states in one shot experiment, if perfect quantum
devices are assumed. The existing protocols amplifying min-entropy sources
\cite{ChungShiWu-PhysicalRandomnessExtractors-2014,
BoudaPawlowskiPivoluskaEtAl-Device-independentrandomnessextraction-2014} use
large number of independent devices, which significantly simplifies the
analysis due to non-existing memory effects. The main upside of this paper is
that we are considering only a single three-partite device to run our
protocol. The price to pay is that our protocol works only for sources with
min-entropy rate $R \geq\frac{1}{4}\log_{2}(10)$.

The paper is organized as follows: In the section II we define the prerequisites
of the protocol. In section III weak random sources are defined and discussed.
The main results are presented for both the zero-error
scenario and risking scenario in section IV, including the discussion on possible
quantum strategies. In section V we conclude and in Appendix we provide the proof
of optimality of the re-send attack.

\section{Prerequisites}

Consider a following scenario: Alice would like to produce perfect random
numbers. She asks her supplier, Eve, to supply a random number generator
(RNG). However, Alice does not really trust in Eve's honesty and would like to
check that Eve really supplied a good RNG and produced bits are random even
conditioned on the knowledge of Eve.

On the other hand Eve would like to influence, or at least learn about the
bits produced by the RNG. To do so, she is granted all power except the
following limitations:

1. Alice's laboratory is safe towards tampering and any communication with
outside world.

2. Alice can ask Eve to deliver RNG in parts. These parts can be prohibited to
communicate within the laboratory among themselves. This can be achieved by perfectly isolating the devices, or by securing space-like separation of the devices during the whole process.

3. Eve is constrained by the laws of quantum mechanics. In particular any
statistics achieved among parts of the RNG must obey relevant Tsirelson's
bounds \cite{Cirel'son-Quantumgeneralizationsof-1980}. Note that it is not enough
to constrain Eve by no-signaling condition as in this case perfect cheating strategies for GHZ game exist.  

4. Alice has a source of somewhat random numbers. Apparently, if Alice has no
such source, Eve could predetermine all steps of Alice in advance, simulate
any results that Alice would expect and use to check honesty of Eve. The level
of randomness of the source needed is a crucial parameter of the protocol and
will be discussed later.

\section{Weak random sources}

We model randomness Alice uses in her protocol by a random variable $X$.
Alice's information about the probability distribution of $X$ is $P(X)$, which
might likely be a perfectly random distribution. We also assume that Eve has a
random variable $E$ with a probability distribution $P(E)$. Eve's information
about $X$ is given by the probability distribution $P(X|E)$ and can be viewed
as the level of correlation between the variables $X$ and $E$. The only
information we suppose about the distribution $P(X|E)$ is that it is random at least to
a certain extent; thus, we allow the output of $X$ conditioned on $E$ to be
distributed according to any probability distribution with sufficient
min-entropy. The goal is then to design an algorithm, which can produce random
outcome independent on the distribution $P(X|E)$.

We say that $X$ conditioned on $E$ contains some randomness if
\begin{equation}
P_{g}(X|E)=\sum_{e}P(E=e)P_{g}(X|E=e)<1,
\end{equation}
where $P_{g}(X|E=e)=\max_{x}P(X=x|E=e)$. This is equivalent to a condition
that for at least one output $e$ that Eve can receive with non-zero
probability, she is unable to predict the output of Alice's random variable
$X$ with certainty.

We quantify the amount of randomness of a distribution by its (conditional)
\emph{min-entropy} defined by
\begin{equation}
H_{\infty}({X}|E)=-\log_{2}P_{g}(X|E).
\end{equation}
Additionally, $X$ is an $(N,k)$ min-entropy source, if it outputs $N$ bit
strings and $H_{\infty}({X}|E) \geq k$. We also define the \emph{min-entropy
rate}
\begin{equation}
\mathbf{R}=\frac{H_{\infty}({X}|E)}{N},
\end{equation}
quantifying bits of entropy of the source per produced random bit.
Min-entropy rate will be used as the figure of merit within this paper,
characterizing the quality of random source used.

Classically it is impossible to extract even a single partially random bit from a single
random source with $N$ bit output and min-entropy smaller than $N-1$. This is
due to the fact that any classical strategy used for the extraction is a
deterministic binary function known by the adversary and she can adjust the
source in such a way that the output of the function is fixed (for details see
e.g.\cite{Shaltiel-IntroductiontoRandomness-2011}).

Here we shortly discuss a different definition of figure of merit of a random
source, namely the so called Santha--Vazirani or SV source introduced in
\cite{SanthaVazirani-Generatingquasi-randomsequences-1986}:

\textit{A string of random bit variables $Z_{i}$ is a $\delta$-SV source
($0\leq\delta\leq\frac{1}{2}$) with respect to $E$ if }
\begin{equation}
\frac{1}{2}-\delta\leq P(Z_{i}=0|E,Z_{1},\dots,Z_{i-1})\leq\frac{1}{2}+\delta.
\end{equation}
For $\delta= 0$ all $Z_{i}$ are uniformly distributed and mutually independent
and any randomness source, even completely deterministic one, can be described
as a SV--source with $\delta= \frac{1}{2}$. Note that all possible
distributions of $N$--bit strings distributed according to a SV-source, have
min-entropy at least $-\log_{2}((\frac{1}{2}+\delta)^{N})$ and the
corresponding min-entropy rate $\mathbf{R}\geq-\log_{2}(\frac{1}{2}+\delta)$.
On the other hand, there are distributions with high min-entropy, which
cannot be characterized as SV--sources with $\delta<\frac{1}{2}$. In
particular a source with a single bit of the sequence fixed and all other bits
perfectly random has min-entropy $H_{\infty}({X}|E)=N-1$ and thus
$\mathbf{R}\rightarrow1$ for large $N$. Nevertheless, it can only be
characterized as a SV--source with $\delta=\frac{1}{2}$.

This is due to the fact that SV--sources assume additional structure of the
randomly distributed $N$--bit strings, namely that the influence of the
adversary is limited locally (per bit), in contrast to the global limitation for
min-entropy sources. This leads to a fact that amplification of SV sources is
much easier in the sense that some amount of randomness is guaranteed to be
present in every single bit from the random source. On the contrary,
min-entropy sources guarantee only the global amount of randomness within the
whole string and any protocol has to be robust especially against sources
which can have a fixed bit value on some of the output bits.

In our work we put only minimal restrictions on the input random source in
terms of min-entropy and due to this fact the proposed procedure can be
utilized on a much broader class of sources comparing to protocols using SV sources.

\section{Amplification of a single weak min-entropy Source}

\begin{figure}[ptb]
\begin{center}
\includegraphics[scale = 5]{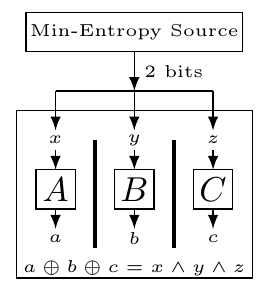} \caption{One round of the protocol.
Two bits from min-entropy source are used to produce inputs into a GHZ test.
The inputs and outputs have to fulfill $a \oplus b \oplus c = x \wedge y \wedge z.$}%
\label{fig1}%
\end{center}
\end{figure}

Our amplification protocol is based on a $GHZ$ paradox \cite{Mermin1990}. The
devices used by the protocol are modeled as three non--communicating black
boxes -- nothing is assumed about their inner workings -- labeled $A,B,C$,
each with two possible inputs $x,y,z\in\{0,1\}$ and two possible outputs
$a,b,c\in\{0,1\}$. One round of the protocol consists of using $2$ bits from a
biased random source $X$ to choose one of the four combinations of inputs
$xyz\in\{111,001,010,100\}$ (see Figure (\ref{fig1})). The round of the protocol
is successful, if the outputs of the boxes fulfill the condition
\begin{equation}
a\oplus b\oplus c=x\wedge y\wedge z, \label{win_con}%
\end{equation}
where $\oplus$ is the logical XOR and $\wedge$ is the logical AND.

It is a well known fact that classical strategies allow the devices to produce
successful outputs with maximum probability $\frac{3}{4}$, while quantum
strategy of measuring the state $|GHZ\rangle=\frac{1}{\sqrt{2}}|000\rangle
+|111\rangle$ in complementary bases ($\sigma_{x}$ if the input is $0$ and
$\sigma_{y}$ if input is $1$) yields the strategy to win with probability $1$.
More importantly, it has been shown in
\cite{ColbeckKent-Privaterandomnessexpansion-2011} that all quantum strategies
succeeding in the protocol with probability $1$ involve measuring (in
complementary bases) GHZ-like states or superpositions of thereof in higher
dimensions. This fact guarantees that if perfect quantum correlations are
observed, the outcomes of the boxes $A$ and $B$ yield two perfectly random
bits and it is exploited in a randomness expansion protocol of
\cite{ColbeckKent-Privaterandomnessexpansion-2011}, where it is supposed that
inputs into black boxes can be chosen with uniform probability.

Here we investigate a scenario, where the inputs for the devices are chosen
according to a source $X$ with conditional min-entropy rate $\mathbf{R}$. The
protocol consists of $n$ rounds, in each round a GHZ paradox is tested. The
necessary number of rounds will be specified later, but generally it depends
on conditional min-entropy rate $\mathbf{R}$ of the source $X$ and desired
quality of the output random bit.

In order to create the inputs into $n$ instances of GHZ test we need to draw
$2n$ bits $X=(R_{1}^{1},R_{2}^{1};R_{1}^{2},R_{2}^{2};\dots;R_{1}^{n}%
,R_{2}^{n})$ from a $(2n,2\mathbf{R}n)$-- weak random source. In the round $i$
two bits $R_{1}^{i}R_{2}^{i}$ are used to choose one out of four possible
input combinations. Let $r = r_{1}^{1}r_{2}^{1},...,r_{1}^{n}r_{2}^{n}$ be a
concrete realization of $X$. Let define $E_{i}(r)$ as
\begin{equation}
E_{i}(r)=-\log_2\left(\max_{kl\in\{0,1\}^{2}} P(R_{1}^{i}R_{2}^{i} = kl \vert R_{1}^{1}R_{2}%
^{2},...,R_{1}^{i-1}R_{2}^{i-1} = r_{1}^{1}r_{2}^{2},...,r_{1}^{i-1}%
r_{2}^{i-1})\right).
\end{equation}

This value characterizes the amount of "fresh" entropy in pair $R_{1}^{i}%
R_{2}^{i}$, given the outcomes of previous rounds. Note that it is possible to
have $E_{i}(r)=0$, so such a source cannot be characterized by a non-trivial
Santha-Vazirani parameter.

Let us label inputs and outputs of $i$-th round as $X_{i},Y_{i},Z_{i}$ (for
example we can set $X_{i}=R_{1}^{i},Y_{i}=R_{2}^{i},Z_{i}=R_{1}^{i}\oplus
R_{2}^{i}\oplus1$) and $A_{i},B_{i},C_{i}$ respectively. Additionally let
$A=(A_{1},\dots,A_{n})$ and $B=(B_{1},\dots,B_{n})$. If in any of the rounds
condition (\ref{win_con}) is not fulfilled, the whole protocol aborts.
Otherwise the outcome of the protocol $O$ is computed as
\begin{equation}
O=Ext(A,B), \label{extractor}%
\end{equation}
where $Ext$ stands for a suitable randomness extractor; its specific choice
will be discussed later.

Let us evaluate the amount of randomness produced by a single round of the
protocol, based on entropy $E_{i}(r)$ present in the two bits $R_{1}^{i}%
R_{2}^{i}$ used for the choice of the inputs. There are two distinct types of rounds:

\begin{enumerate}
\item It holds that $E_{i}(r)>\log_{2}(3)$; in this case $P(R_{1}^{i}R_{2}%
^{i}=x_{i}y_{i}|R_{1}^{1}R_{2}^{1},...,R_{1}^{j}R_{2}^{j}=r_{1}^{1}r_{2}%
^{1},...,r_{1}^{j}r_{2}^{j})>0$ for all four possible values of $x_{i}%
,y_{i}\in\{00,01,10,11\}$ and the only strategy fulfilling condition
(\ref{win_con}) with probability $1$ is the honest strategy of measuring GHZ
states. As discussed before, in this case bits $A_{i}$ and $B_{i}$
are uniformly distributed and independent of each other as well as all the
other previous inputs $X_{j},Y_{j},Z_{j},\quad j\in\{1,\dots,i\}$ and outputs
$A_{j},B_{j},C_{j},\quad j\in\{1,\dots,i-1\}$. Probability of any other
strategy to fulfill (\ref{win_con}) is bounded away from $1$ as proven in
\cite{ColbeckKent-Privaterandomnessexpansion-2011}.

\item It holds that $E_{i}(r)\leq\log_{2}(3)$; there exists a probability
distribution $P$, such that $P(R_{1}^{i}R_{2}^{i}=x_{i}y_{i}|R_{1}^{1}%
R_{2}^{2},...,R_{1}^{i-1}R_{2}^{i-1}=r_{1}^{1}r_{2}^{2},...,r_{1}^{i-1}%
r_{2}^{i-1})=0$ for at least one possible value of $x_{i},y_{i}\in
\{00,01,10,11\}$. In this case there exists a classical strategy (which can be
encoded in the common information $\lambda$) that fulfills condition
(\ref{win_con}) with probability $1$.
\end{enumerate}

Based on this preliminary analysis, it is clear that for $\mathbf{R}\leq
\log_{2}(\sqrt{3})$ there exists a probability distribution on $2n$ bit
strings, such that each of $n$ rounds of the protocol is of type 2, and there
is a classical deterministic strategy to win all of them.

However, if $\mathbf{R}>\log_{2}(\sqrt{3})$, with some non-zero
probability there will be some rounds of type 1 during the run of the protocol.

\subsection{Zero error protocol}

Let us first consider a scenario, in which the adversary doesn't want to risk
getting caught at all. In this case rounds of type 1 need to be played with
honest GHZ strategy. This fact alone however doesn't guarantee that the output
of the protocol contains some non-zero amount of randomness, because the
rounds of type 2 can in principle depend on the outcomes of the earlier rounds
of type 1.

As an example consider a very general scenario where the resulting bit $O$ is computed
as a sum of partial results from individual rounds $o_i$. Let $o_i$ be a result of a round $i$ of type 1,
arbitrarily random. Let $j$ by a subsequent round of type 2. Devices and source can agree in advance that
in round $j$ they will output results obtained in the round $i$ independently on the inputs. In such a case
$o_j = o_i$ and $o_j \oplus o_i =0$ and thus perfectly deterministic. The win condition  (\ref{win_con}) will
also be automatically satisfied. The price to pay is the fact that the source had to select a specific
outcome in the round $j$, which decreases its entropy.

In what follows, we will analyze to what extent can the outcomes of the rounds
of type 2 negate any randomness produced in the rounds of type 1, given a specific entropy of the source. Assume that
$k$ out of $n$ rounds are of type 1. Without the loss of generality we can
assume that all $k$ rounds of type 1 are realized before $n-k$ rounds of type
2. In fact, this order of rounds gives the adversary the best possible
situation to react in rounds of type 2 on the randomness already produced in
rounds of type 1.

In such ordering we have:
\begin{align}
A  &  =\left(  \vec{A}_{k},f_{\lambda}^{k+1}(\vec{A}_{k}),\dots,f_{\lambda
}^{n}(\vec{A}_{k})\right)  ,\nonumber\\
B  &  =\left(  \vec{B}_{k},g_{\lambda}^{k+1}(\vec{B}_{k}),\dots,g_{\lambda
}^{n}(\vec{B}_{k})\right)  ,
\end{align}
where $\vec{A}_{k}=(A_{1},\dots,A_{k}),\vec{B}_{k}=(B_{1},\dots,B_{k})$ are
outcomes of the rounds of type 1. Functions $f_{\lambda}^{j}$ and $g_{\lambda
}^{j}$, $k+1\leq j\leq n$ are particular strategies in round $j$ of type 2
attempting to increase the bias of the final bit, depending on the outcomes of
the rounds of type 1 and common information $\lambda$. Recall that $\lambda$
is the common information between the devices, source and the adversary. All
these parties can be correlated only via this random variable. In a regime
where the adversary doesn't want to risk getting caught at all, this means
that although vectors $A$ and $B$ are generally not independent, they can only
be dependent via $\lambda$. Therefore given $\lambda$, $A$ and $B$ are
independent and their respective conditional min-entropies are $H_{\infty
}(A|\lambda)=H_{\infty}(B|\lambda)=k$. Thus we can use any two source
extractor $Ext$ to extract the entropy present in $A$ and $B$. Since $A$ and
$B$ are independent given $\lambda$, it holds that $\left(  Ext(A,B)|\lambda
\right)  $ will be distributed according to the properties of the particular
extractor (close to being uniformly distributed given $\lambda$).

\subsection{Hadamard extractor}

In our analysis we choose a particular form of the extractor (\ref{extractor}%
), which defines our output bit as
\begin{equation}
O=Had(A,B)=\bigoplus_{i=1}^{n}(A_{i}\wedge B_{i}).
\end{equation}
This extractor is called Hadamard extractor in the literature
\cite{ChorGoldreich-UnbiasedBitsfrom-1988,
BoudaPivoluskaPlesch-ImprovingHadamardExtractor-2012,
DodisElbazOliveiraEtAl-ImprovedRandomnessExtraction-2004} and it guarantees
that $\left(  Ext(A,B)|\lambda\right)  $ is $2^{(n-H_{\infty}(A|\lambda
)-H_{\infty}(B|\lambda)-2)/2}$-close to a uniformly distributed bit as long as
$H_{\infty}(A|\lambda)+H_{\infty}(B|\lambda)\geq\frac{n}{2}$. Therefore as
long as $k>\frac{n}{2}$, regardless of the strategy employed in rounds of type
2, the output bit is, at least to some extent, random. Note here that the
requirement on $k$ could in principle be made lower by using different two-source extractors.
For example Bourgain's extractor
\cite{BOURGAIN-MORESUM-PRODUCTPHENOMENON-2005} produces non-deterministic bit
as long as the sum of the entropies of $A$ and $B$ is greater than
$2n(1/2-\alpha)$ for some universal constant $\alpha$ and non-explicit
extractors can go as low as $k=O(\log n)$
\cite{ChorGoldreich-UnbiasedBitsfrom-1988}.

In the light of the previous analysis we can obtain the upper bound for the
min-entropy rate, for which full cheating (maximum bias with probability of
getting caught equal to $0$) is possible. In order to do so, let us represent
$2n$ bit strings that the biased source $X$ can output with non-zero
probability by a graph tree of depth $n$, where

\begin{itemize}
\item each of the vertices has at most $4$ children and each edge from parent
to child is labeled by one of $\{00,01,10,11\}$,

\item each vertex represents prefix of a concrete realization of $r$ with
$r_{1}^{1},r_{2}^{1},\dots,r_{1}^{i},r_{2}^{i}$ encoded in the edge labels on
the path from the root of the tree to the given vertex,

\item each leaf represents a concrete realization of $r$.
\end{itemize}

Clearly, each vertex has at least $\left\lceil 2^{E_{i}(r)}\right\rceil$ children. A vertex
with $2^{E_{i}(r)}>3$ will be called an honest vertex, as in this vertex an honest
- quantum strategy must be used, whereas all other vertices will be called
dishonest vertices.

\begin{figure}[ptb]
\begin{center}
\includegraphics[scale = 2]{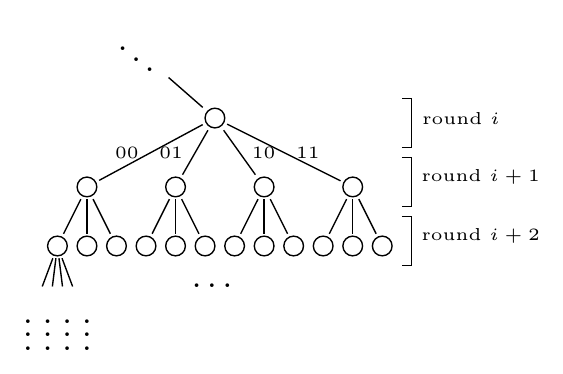} \caption{Optimal tree
representation of the random variable that potentially enables full cheating alternates
between honest and dishonest vertices.}%
\label{randomtree}%
\end{center}
\end{figure}

To give an upper bound on the min-entropy for which the adversary can fully
cheat, we need to find a tree with a maximal number of leafs, such that for
each path from the root to the leaf the number of honest vertices is smaller
or equal to the number of dishonest vertices. Apparently such tree can be
constructed by alternating between honest and deterministic vertices along
each path (see Figure (\ref{randomtree})); such tree has $\sqrt{12}^{n}$
leafs. Uniform distribution over $\sqrt{12}^{n}$ leaves maximizes the
min-entropy that can be used to realize such tree, yielding the min entropy
rate of $\mathbf{{R}}_{H}\mathbf{=}\log_{2}\mathbf{(}12\mathbf{)/}4$. For any
higher min-entropy rate, there exists a leaf such that the number of honest
vertices on the path from the root to the leaf is higher than the number of
dishonest vertices, therefore the adversary cannot know the outcome of the
protocol with probability $1$ without risking to be caught.

If the actual min-entropy rate of the source used is expressed as $\mathbf{R=R}%
_{H}+\varepsilon$ with arbitrary $\varepsilon>0$, the probability of every
single leaf in the tree will be upper bounded by $p_{1}=2^{-2n\mathbf{R}%
}=\sqrt{12}^{-n}2^{-2\varepsilon n}$. In such a tree no more than $\sqrt
{12}^{n}$ leaves will be of a form that allows cheating without risking to be
caught, so the overall probability of cheating success is bounded from above by
\begin{equation}
p_{cheat}\leq2^{-2\varepsilon n},\label{p_cheat}%
\end{equation}
thus decreasing to zero exponentially with $n$. With this probability a bias of the output bit $\frac{1}{2}$
is achieved, whereas in all other leafs the bias is $0$, so the resulting bias
of the output bit will be
\begin{equation}
bias(B)\leq2^{-(2\varepsilon n+1)}.\label{bias}%
\end{equation}

\begin{figure}[ptb]
\begin{center}
\includegraphics[scale = 2]{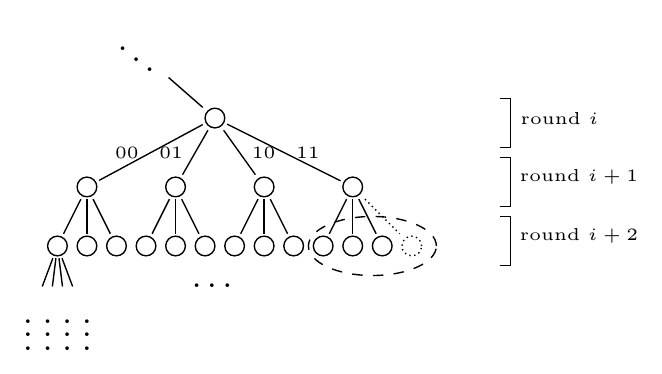} \caption{By adding a fourth
child to a dishonest vertex, it is converted to an honest one in the zero-error scenario. Thus all its children
(in the dashed oval)
have positive probability to appear - therefore they all produce random outcomes in zero-error scenario.
In the risking scenario, the vertex stays dishonest,
the added (dotted) child leads to abortion of the protocol, however the other children remain deterministic.
}%
\label{randomtree_3}%
\end{center}
\end{figure}

It is worth to mention that
with growing $\mathbf{R}$ the number of cheatable leaves is in fact decreasing
and the actual cheating probability and consequently also the resulting bias will thus be strictly lower. This is due
to the fact that with every extra leave added to the probability tree, some
other leaves will convert from a fully biased to a perfectly random outcome.
This is due to the fact that the extra leaves can be added only by adding a
fourth child to a dishonest vertex, which is in this way converted to an
honest one, resulting into honestness of its leafs (for depiction see Figure
(\ref{randomtree_3})).

\begin{figure}[ptb]
\begin{center}
\includegraphics[scale = 2]{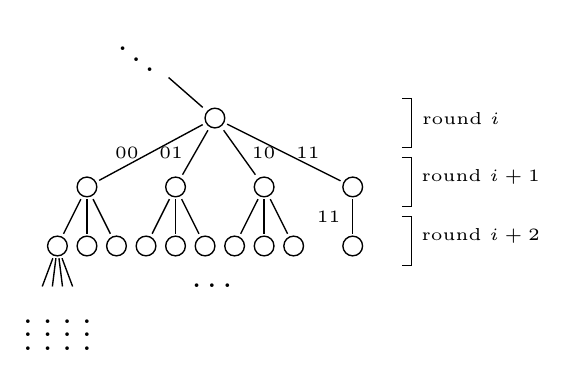}\caption{The optimal achievable strategy is
to repeat outcomes of the previous honest round. Random inputs
$\{001,010,100\}$ are interchangeable, while input $\{111\}$ needs to be
exactly repeated in the dishonest round.}%
\label{randomtree2}%
\end{center}
\end{figure}

The rate $\mathbf{R}_{H}=\log_{2}(12)/4$ is only an upper bound for the amount
of min-entropy for which the full cheating is possible. In fact, there is no
constructive attack that would be possible with such a min-entropy rate. As
shown in the appendix, the optimal implementable strategy is the one mentioned
earlier -- in every other round the boxes simple resend the outcomes of the
previous honest round. Such strategy can tolerate less min-entropy than
$\mathbf{R}_{H}$, as the dishonest vertex connected to it's honest parent by a
$11$ edge must have only one child, also labeled $11$ (see Figure
\ref{randomtree2}).

Uniform distribution over the leaves of such tree has a
min-entropy rate
\begin{equation}
\mathbf{R}_{\max}=1/4\log_{2}10,
\end{equation}
which is the highest rate for which full cheating is possible -- half of the
rounds are of type 1, quantum and honest, and half of the rounds are
of type 2, negating the bias of the output obtained of the previous runs. As soon as $\mathbf{R}%
=\mathbf{R}_{\max}+\epsilon$, the resulting bias
exponentially converges to zero with the same arguments as used for Hadamard extractor.

\subsection{Risking to fail}

In a realistic scenario, if it would not be possible for Eve to limit the
inputs as needed for full cheating (i.~e. $\mathbf{R}>\log_{2}(12)/4$), Eve
could simply try to use a classical strategy and guess the correct outcomes in
some of the honest rounds. Let us now analyze, what would be the probability
of successful cheating with such a strategy.

One can model such cheating strategy by adding extra leaves to the fully
cheatable tree (See Figure (\ref{randomtree_3})). This can be achieved by adding a fourth edge to some
of the dishonest vertices. In such round, Eve would simply use a classical
strategy, which is successful only in three out of four realizations.
Therefore, if this new added edge is actually realized by the random source, the
protocol fails by not satisfying (\ref{win_con}). The number of leaves for
which this strategy is successful stays exactly $\sqrt{12}^{n}$ and all the
other leaves lead to failure of the protocol. With a min-entropy rate
$\mathbf{R=R}_{H}+\varepsilon$ the minimal number of leaves in the tree is
$\sqrt{12}^{n}2^{2\varepsilon n}$, thus the probability of not failing the
protocol is $p_{guess}=2^{-2\varepsilon n}$.
Comparing to $p_{cheat}$ (\ref{p_cheat}) we see that the probability of successfully
cheating the protocol by risking is the same as the upper bound
of the probability of successful cheating of the protocol without risking.

One might think about a more general attack where some reasonable bias is
achieved with a very small probability of the protocol to fail by using a
quantum strategy utilizing other than GHZ states. We limited ourselves to the
analysis of quantum strategies that do not use entangled states across the
individual rounds of the protocol. Using the SDP introduced in
\cite{MironowiczPawlowski-Amplificationofarbitrarily-2013} we numerically showed that the bias
of the output bit $b$ achievable by any quantum strategy, if the
failure probability in every round is upper bounded by $p_{f}$, is upper bounded by $bias(b)^{2}\leq
p_{f}$. Such strategy can be in fact realized
by using states close to GHZ and by suitable changing the measurements used by
the devices to POVM measurements with some pre-shared classical information.
By utilizing this approach the adversary can slightly adjust results of the
non-corrected quantum rounds with only a small probability of aborting the
protocol. But due to the polynomial dependence between the bias and the
failure probability, either the bias of the output bit or the probability of
successful finishing of the protocol stays exponentially small.

\section{Conclusion}

We have presented a scheme for production of almost perfect random numbers
with intrusted quantum devices and a single weak random source, based on
tri-partite GHZ-type entanglement. This scheme can be used for production of
perfect randomness by parallel monitoring of the honesty of the devices. We
use min-entropy to characterize the randomness of the source, which
guarantees the minimal possible assumptions about its detail characteristics,
in particular about any local behavior. This allows amplification of a wide
variety of weak random sources including sources not amplifiable by already
known procedures.

In contrast to previous results, here we repeatedly reuse the same devices
during the amplification protocol, using just three independent devices with
arbitrary amount of memory available. A drawback of the protocol, left for the future research, is that
it is not fault-tolerant -- it will abort with the first wrong output and
thus is not experimentally feasible yet.

\textit{Acknowledgement --} We acknowledge the support of the Czech Science
Foundation GA\v{C}R project P202/12/1142, EU project RAQUEL, as well as project VEGA 2/0072/12.


\appendix

\section{Optimality of the "Re-send" strategy}

Let us here analyze, if and to what extent can the adversary, using protocol
rounds of type 2, bias the output bit $B$. To effectively do this, the
strategy for output in these rounds must depend on outputs of some of the
previous rounds - any fixed strategy would produce only fixed bits that would
not influence the bias of the final bit $B$. We show, that the optimal strategy in round of type
2 is simply to resend outputs created by honest strategy in a single previous
round of type 1.

In the round $j$, which is considered to be of the type 2, the outputs of the
devices $A$, $B$, and $C$ will be respectively%
\begin{align}
&  f_{\lambda}^{j} (a_{1},\dots,a_{j-1},x_{1},\dots,x_{j})\\
&  g_{\lambda}^{j}(b_{1},\dots,b_{j-1},y_{1},\dots,y_{j})\nonumber\\
&  h_{\lambda}^{j}(c_{1},\dots,c_{j-1},z_{1},\dots,z_{j}),\nonumber
\end{align}
where we assume that all rounds $1,\dots,j-1$ are honest. This assumption
doesn't change the generality of the result, because previous rounds of type 2
are deterministic and do not add any randomness into the final output $B$.

Outcomes of these functions also need to fulfill the condition (\ref{win_con}%
), that can be considered in the form%
\begin{equation}
f_{\lambda}^{j}\oplus g_{\lambda}^{j}\oplus h_{\lambda}^{j}=x_{j}\wedge
y_{j}\wedge z_{j}(\lambda,x_{1},y_{1},z_{1},...,x_{j-1},y_{j-1},z_{j-1}),
\label{win_con_2}%
\end{equation}
where $x$ and $y$ are parameterized the same way as $z$. At the first glance
it might look like there is no way that functions $f$, $g$ or $h$ might depend
on previous outputs, as the right hand side of (\ref{win_con_2}) cannot depend
on it. This is however not entirely true. The fact that the protocol continues
means that conditions (\ref{win_con}) for all previous rounds were fulfilled
and the condition (\ref{win_con_2}) can be thus rewritten (omitting explicit
dependence on the parameters $x$, $y$ and $z$) as%
\begin{equation}
f_{\lambda}^{j}\oplus g_{\lambda}^{j}\oplus h_{\lambda}^{j}=x_{j}\wedge
y_{j}\wedge z_{j}(\lambda,a_{1}\oplus b_{1}\oplus c_{1},...,a_{j-1}\oplus
b_{j-1}\oplus c_{j-1}). \label{win_con_3}%
\end{equation}

Let us now examine the condition (\ref{win_con_3}) in more detail. The goal is
to show that functions $f_{\lambda}^{j}$, $g_{\lambda}^{j}$, and $h_{\lambda}^{j}$ have
to have a specific form, depending only on common information $\lambda$. To
demonstrate this, consider an arbitrary (honest) round $i < j$. Let us define
vectors of parameters $\vec{x}_{j} = (x_{1},\dots,x_{j-1})$, $\vec{a}_{i} =
(a_{1},\dots,a_{i-1},a_{i+1},\dots, a_{j-1})$ and partial function
$f^{j}_{\lambda, \vec{a}_{i}, \vec{x}_{j}}(a_{i}) = f_{\lambda}^{j}%
(a_{1},\dots,a_{j-1},x_{1},\dots,x_{j})$. Also consider $g^{j}_{\lambda,
\vec{b}_{i}, \vec{y}_{j}}(b_{i})$ and $h^{j}_{\lambda, \vec{c}_{i}, \vec
{z}_{j}}(c_{i})$ defined analogously. All $f^{j}_{\lambda, \vec{a}_{i},
\vec{x}_{j}}(a_{i})$, $g^{j}_{\lambda, \vec{b}_{i}, \vec{y}_{j}}(b_{i})$ and
 $h^{j}_{\lambda, \vec{c}_{i}, \vec{z}_{j}}(c_{i})$ are functions mapping one
bit into one bit. There are only four functions of this type, two are
\emph{constant} -- mapping both inputs into a constant output $0$ or $1$, and
two of them are \emph{balanced} -- identity and negation.

Now we can rewrite (\ref{win_con_3}) as
\begin{align}
& f^{j}_{\lambda, \vec{a}_{i}, \vec{x}_{j}}(a_{i}) \oplus g^{j}_{\lambda,
\vec{b}_{i}, \vec{y}_{j}}(b_{i}) \oplus h^{j}_{\lambda, \vec{c}_{i}, \vec
{z}_{j}}(c_{i}) =\nonumber\\
& = x_{j}\wedge y_{j}\wedge z_{j}(\lambda,a_{1}\oplus b_{1}\oplus
c_{1},...,a_{j-1}\oplus b_{j-1}\oplus c_{j-1}).\label{win_con_4}%
\end{align}

In the next step let us grant the source additional knowledge. Namely, let us
suppose that the inputs $x_{j},y_{j},z_{j}$ for $j^{th}$ round are chosen with
the full knowledge of outputs of rounds $1,\dots,i-1,i+1,\dots,j-1$, along
with all the previous inputs. \ This is equivalent to the full knowledge of
functions $f^{j}_{\lambda, \vec{a}_{i}, \vec{x}_{j}}() , g^{j}_{\lambda,
\vec{b}_{i}, \vec{y}_{j}}(), h^{j}_{\lambda, \vec{c}_{i}, \vec{z}_{j}}()$.

Therefore the only knowledge the source doesn't posses about the outcomes of
the $j$th round are values of $a_{i},b_{i}$ and $c_{i}$. However, it knows
their $XOR$ $a_{i}\oplus b_{i}\oplus c_{i}$. This knowledge allows it to
discriminate between quadruples of possible outputs. They were either from
$\{000,110,101,011\}$, if $a_{i}\oplus b_{i}\oplus c_{i} = 0$, or from
$\{111,001,010,100\}$, if $a_{i}\oplus b_{i}\oplus c_{i} = 1$. In order to
fulfill (\ref{win_con_4}), $f^{j}_{\lambda, \vec{a}_{i}, \vec{x}_{j}}(a_{i}) ,
g^{j}_{\lambda, \vec{b}_{i}, \vec{y}_{j}}(b_{i}), h^{j}_{\lambda, \vec{c}_{i},
\vec{z}_{j}}(c_{i})$ must append the same output for all $a_{i}$, $b_{i}$ and $c_{i}$
with the same XOR.



Now we show that to fulfill (\ref{win_con_4}), all $f^{j}_{\lambda, \vec
{a}_{i}, \vec{x}_{j}}(a_{i})$, $g^{j}_{\lambda, \vec{b}_{i}, \vec{y}_{j}}%
(b_{i})$ and $h^{j}_{\lambda, \vec{c}_{i}, \vec{z}_{j}}(c_{i})$ must be either
simultaneously \emph{constant} or simultaneously \emph{balanced}. We proceed
with a proof by contradiction and without loss of generality assume
$f^{j}_{\lambda, \vec{a}_{i}, \vec{x}_{j}}(a_{i})$ is constant and
$g^{j}_{\lambda, \vec{b}_{i}, \vec{y}_{j}}(b_{i})$ is balanced. Then there is
no function $h^{j}_{\lambda, \vec{c}_{i}, \vec{z}_{j}}(c_{i})$ which can
fulfill the condition (\ref{win_con_4}) (see table (\ref{table1})).

\begin{table}[h]%
\begin{tabular}
[c]{|c|c|c|c|c|}\hline
$a_{i}\oplus b_{i} \oplus c_{i}$ & $a_{i}b_{i}c_{i}$ & $f^{j}_{\lambda,
\vec{a}_{i}, \vec{x}_{j}}(a_{i}) = 0$ & $g^{j}_{\lambda, \vec{b}_{i}, \vec
{y}_{j}}(b_{i}) = id$ & ${h}^{j}_{\lambda, \vec{c}_{i}, \vec{z}_{j}}(c_{i}%
)$\\\hline
1 & 001 & 0 & 0 & 1/0\\
1 & 010 & 0 & 1 & 0/1\\
1 & 100 & 0 & 0 & 1/0\\
1 & 111 & 0 & 1 & 0/1\\\hline
0 & 011 & 0 & 1 & 1/0\\
0 & 101 & 0 & 0 & 0/1\\
0 & 110 & 0 & 1 & 1/0\\
0 & 000 & 0 & 0 & 0/1\\\hline
\end{tabular}
\caption{Without loss of generality we chose $f^{j}_{\lambda, \vec{a}_{i},
\vec{x}_{j}}(a_{i}) = 0$ and $g^{j}_{\lambda, \vec{b}_{i}, \vec{y}_{j}}%
(b_{i})$ identity. Choosing any other combination of constant/balanced
function will result only in negation of the appropriate column. Since the
source has only a limited information about outcomes $a_{i},b_{i},c_{i}$,
namely their XOR, it can discriminate between quadruples of inputs in second
column. For this reason $f^{j}_{\lambda, \vec{a}_{i}, \vec{x}_{j}}(a_{i})
\oplus g^{j}_{\lambda, \vec{b}_{i}, \vec{y}_{j}}(b_{i}) \oplus h^{j}_{\lambda,
\vec{c}_{i}, \vec{z}_{j}}(c_{i})$ must be constant for these quadruples and we
filled in column of $h^{j}_{\lambda, \vec{c}_{i}, \vec{z}_{j}}(c_{i})$
accordingly. However, such column doesn't define a function, which is a
contradiction. }%
\label{table1}%
\end{table}

It remains to show that the constant/balanced property can depend only on
previously shared information $\lambda$. To show this is indeed true, let us
fix the value of vectors $\vec{a}_{i}$ and $\vec{x}_{i}$ and without loss of
generality suppose $f^{j}_{\lambda, \vec{a}_{i}, \vec{x}_{j}}(a_{i})$ is
balanced. Because all rounds $i<j$ are honest, each value of vectors $\vec
{b}_{i}$ and $\vec{y}_{i}$ can appear in a run with $\vec{a}_{i}$ and $\vec
{x}_{i}$ with non-zero probability. Therefore, in order to fulfill
(\ref{win_con_4}) with probability 1, $g^{j}_{\lambda, \vec{b}_{i}, \vec
{y}_{j}}(b_{i})$ must be balanced for each value of $\vec{b}_{i}$ and $\vec
{y}_{i}$. By symmetry, the same argument can be used to show that the
constant/balanced property can depend only on $\lambda$.

Without the loss of generality we can suppose that part of the common
information $\lambda$ contains information about the partial function for
each $i$ and $j$ in the form of $(\alpha_{i}^{j},\lambda_i)$.
The partial functions are fully specified by these parameters as%

\begin{align}
f^{j}_{\lambda, \vec{a}_{i}, \vec{x}_{j}}(a_{i})  & = f^{j}_{\lambda}(a_{i}) =
(\alpha_{i}^{j} \wedge a_{i}) \oplus f_{\lambda_{i}}^{j}(\vec{a}_{i},\vec{x}%
_{j})\\
g^{j}_{\lambda, \vec{b}_{i}, \vec{y}_{j}}(b_{i})  & = g^{j}_{\lambda}(a_{i}) =
(\alpha_{i}^{j} \wedge b_{i}) \oplus g_{\lambda_{i}}^{j}(\vec{b}_{i},\vec
{y}_{j})\nonumber\\
h^{j}_{\lambda, \vec{c}_{i}, \vec{z}_{j}}(c_{i})  & = h^{j}_{\lambda}(a_{i}) =
(\alpha_{i}^{j} \wedge c_{i}) \oplus h_{\lambda_{i}}^{j}(\vec{c}_{i},\vec
{z}_{j}).\nonumber
\end{align}

The parameter $\alpha_{i}^{j}$ specifies if the partial function in $j^{th}$
round for $i^{th}$ parameter is constant ($\alpha_{i}^{j} = 0$) or balanced
($\alpha_{i}^{j} = 1$), and parameter $\lambda_{i}$
specifies concrete functions of previous inputs and outputs. These functions can
only influence which concrete balanced or concrete constant function is used
in the given round (if the function is equal to 1 XOR effectively negates the output).
Recall that this needs
to hold for all $i<j$. If we define $S_{j} = \{i\vert i<j,
\alpha_{i}^{j} = 1\}$, we have the following form of the functions%

\begin{align}
&  f_{\lambda}^{j} (a_{1},\dots,a_{j-1},x_{1},\dots,x_{j})= \bigoplus
\limits_{i\in S_{j}}a_{i}\oplus f_{j}^{\prime}\nonumber\\
&  g_{\lambda}^{j}(b_{1},\dots,b_{j-1},y_{1},\dots,y_{j})= \bigoplus
\limits_{i\in S_{j}}b_{i}\oplus g_{j}^{\prime}\nonumber\\
&  h_{\lambda}^{j}(c_{1},\dots,c_{j-1},z_{1},\dots,z_{j})= \bigoplus
\limits_{i\in S_{j}}c_{i}\oplus h_{j}^{\prime},
\end{align}
where $f_{j}^{\prime},g_{j}^{\prime},h_{j}^{\prime}$ depend only on common
information $\lambda$, inputs into concrete devices and outputs of rounds $\ell\notin
S_{j}$, thus effectively only choosing between negation of the output or identity.

Let us now analyze to what extent these outputs in a specific round can help
to bias the final output bit. With $k$ rounds of the type 1 and a single round
of type 2 the output bit will have the form (up to a constant factor not
changing the bias)%
\begin{equation}
B=\bigoplus\limits_{i=1}^{k}\left(  a_{i}\wedge b_{i}\right)  \oplus\left(
\bigoplus\limits_{i\in S}a_{i}\wedge\bigoplus\limits_{i\in S}b_{i}\right)  .
\label{B}%
\end{equation}
If $S$ is empty, trivially the output bias is ${2^{-(k+1)}}$. If $S$ has one
element, say $t$, with the $k+1$st round the result of $t$-th round is
repeated (thus effectively negating it) and the bias of the output result
decreases to ${2^{-k}}$. To evaluate the output for $s =\vert S \vert> 1$, let
us rewrite the expression (\ref{B}) into the form%
\begin{equation}
B=\bigoplus\limits_{i\notin S}\left(  a_{i}\wedge b_{i}\right)  \oplus\left[
\bigoplus\limits_{i\in S}\left(  a_{i}\wedge b_{i}\right)  \oplus\left(
\bigoplus\limits_{i\in S}a_{i}\wedge\bigoplus\limits_{i\in S}b_{i}\right)
\right] .
\end{equation}
The first part of the expression depends only on the rounds not incorporated
in the set $S$ and is easy to compute; it yields a bias $2^{-(k+1-s)}$.

Let us now calculate the bias of the second part. Denote as $k_{a}$ and
$k_{b}$ the number of $1$'s in all outputs $a_{i}$ and $b_{j}$ with $i,j\leq
s$ respectively. It is easy to see that the correcting round will change the
output bit if and only if both $k_{a}$ and $k_{b}$ are odd. We can now
calculate the bias as the difference between the fraction of outputs $0$ and
$1/2$:%
\begin{align}
& bias\left(  B\right)  =\nonumber\\
& = \left\vert \frac{1}{2^{2s}}\sum_{k_{a},k_{b}=0}^{s}\binom{s}{k_{a}}%
\sum_{i=\min}^{\max}\binom{k_{a}}{i}\binom{s-k_{a}}{k_{b}-i}\frac{1+\left(
-1\right)  ^{i+k_{a}k_{b}}}{2} - \frac{1}{2}\right\vert \nonumber\\
& \min=Max\left(  0,k_{a}+k_{b}-s\right) \\
& \max=Min\left(  k_{a},k_{b}\right)  .\nonumber
\end{align}
This sum can be evaluated and yields $2^{-(s+1)}$ for $s$ even and $2^{-s}$
for $s$ odd. Thus, altogether the bias of the output bit is either
$2^{-(k+1)}$ if $s$ is even or $2^{-k}$ for $s$ odd. So we can conclude that
the best result the adversary can hope for with one round of type 2 is to
negate the result of one of the previous rounds of the type 1.

\end{document}